\documentclass{elsart3}

\usepackage{amssymb}
\usepackage{graphicx}
\journal{Physica C}
\begin{document}

\begin{frontmatter}



\title{Microwave properties of
YBa$_2$Cu$_3$O$_{7-\delta}$ films with BaZrO$_3$ nanoinclusions.}

\author[tre]{N. Pompeo\corauthref{cor}},
\corauth[cor]{Corresponding author} \ead{pompeo@fis.uniroma3.it}
\author[tre]{E. Silva},
\author[tre]{R. Marcon},
\author[enea]{V. Galluzzi},
\author[enea]{G. Celentano},
\author[enea]{A. Augieri},
\author[tun]{T. Petrisor},

\address[tre]{Dipartimento di Fisica ``E.Amaldi'' and
Unit\`{a} CNISM, Universit\`{a} di Roma Tre, V. Vasca Navale 84, 00146
Roma, Italy}
\address[enea]{ENEA, 00044 Frascati, Roma, Italy}
\address[tun]{Technical University, 3400 Cluji-Napoca, Romania}

\begin{abstract}
We present measurements of the microwave complex surface impedance at
47.7 GHz in YBa$_2$Cu$_3$O$_{7-\delta}$ (YBCO) films deposited by pulsed
laser deposition with the explicit goal to introduce BaZrO$_3$ (BZO)
nanoinclusions.  Composite targets obtained by addition of
BZO powder in molar percents ranging from 2.5 to 7 mol.\% have been
prepared and characterized.  Measurements of the microwave surface
impedance indicate a broadened transition in zero field, however
compensated by a very large pinning frequency, indicating that while
intergrain properties are still to be optimized the effect of
nanometric inclusions substantially enhances the intragrain vortex
pinning.
\end{abstract}

\end{frontmatter}


Various techniques have been exploited in order to improve vortex
pinning and reduce vortex motion. In films both intergrain (grain
boundary dominated, GB) and intragrain phenomena play a role in
determining the magnetic field-dependent dissipation properties.
While GB dissipation dominates at low magnetic fields, the intra-grain
mechanisms constitute the ultimate limitation.  Film texture is
beneficial for reducing GB dissipation, while the introduction of
artificial defects of order of the vortex diameter ($\sim$ the
coherence length) can be effective in order to increase vortex pinning
and reduce vortex motion.  In the attempt of achieving high-$J_{c}$
YBCO coated conductor a renewed interest arose about the investigation
of novel or improved techniques for the introduction of pinning
centers.
A possible, feasible method to introduce such defects is the inclusion
of nanometer-size particles in the target used for film deposition 
\cite{mcmanus}.
\begin{figure}[htb]
\includegraphics [width=7.3cm]{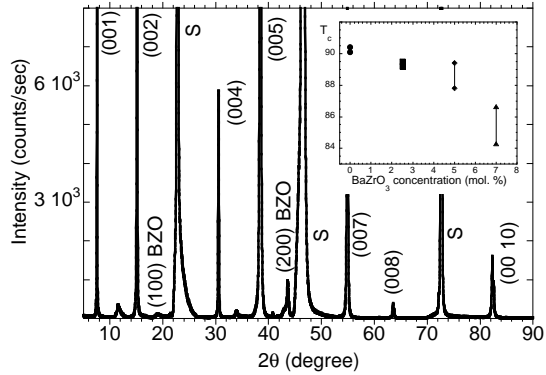} \caption{X-ray
$\theta-2\theta$ scan. Inset: $T_{c}$ vs. BZO content.}
\label{fig1}
\end{figure}
We have grown YBCO films (thicknes $d\sim$200 nm) with 
BZO inclusions by pulsed
laser ablation \cite{fabbri} starting from mixed YBCO-BZO sintered
targets with various BZO concentration ranging from 7 to 2.5 percent,
on single crystal SrTiO$_{3}$ and CeO$_{2}$ buffered $r$-cut
alpha Al$_{2}$O$_{3}$ (for  microwave 
measurements) substrates.
The dc zero resistance critical temperature $T_{c}$
revealed a slight dependence on the BZO content
(inset of Fig.\ref{fig1}).  We focus here on the results obtained on
the 7\% samples. Only (00l) YBCO and substrate (S) peaks are visible in the $\theta-2\theta$ scan (Fig. \ref{fig1}),
indicating a good epitaxy.  $\omega$ scan through the (005) YBCO peak
exhibits FWHM of 0.1$^{\circ}$.  Other reflections at about
21.5$^{\circ}$ and 43.7$^{\circ}$ are present, ascribable to (h00)
BZO.\\
The measurement of the microwave response in the vortex state is a
particularly stringent test of the pinning properties, since small
vortex oscillations are involved and one tests both the
presence of pinning centers and the steepness
of the potential well. Not
too close to $T_{c}$, and when the field dependence of the complex 
resistivity is linear, the simplest models for fluxon motion
\cite{grcc} can be applied, and one has:
\begin{equation}
\label{ro}
\frac{\Delta\rho_{1}+\rm{i}\Delta\rho_{2}}{\Delta B}=
\frac{\Phi_{0}}{\eta}
\frac{1+\mathrm{i}\frac{\nu_{p}}{\nu}}{1+\left(\frac{\nu_{p}}{\nu}\right)^{2}}
\end{equation}
where $\Delta\rho$ is the complex resistivity change due to a field
variation $\Delta B$, $\eta$ is the fluxon viscosity and the pinning
frequency $\nu_{p}$ ($\sim$10 GHz above 70 K in YBCO crystals \cite{tsuchiya})
is a measure of the steepness of the potential well for the flux
lines.  \\
\begin{figure}[htb]
\includegraphics [width=7.0cm]{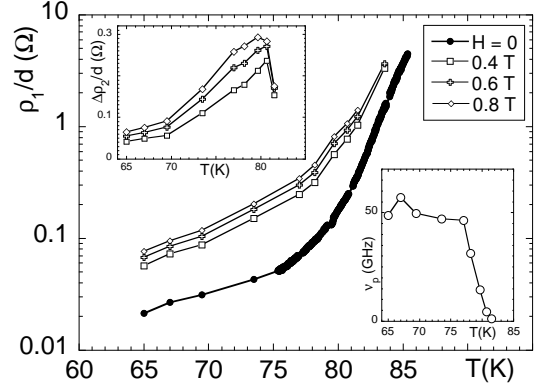} \caption{Microwave complex
resistivity at 47.7 GHz at different fields.  Main panel: $\rho_{1}/d$.
Upper inset:
$\Delta\rho_{2}/d=\left[\rho_{2}(H,T)-\rho_{2}(0,T)\right]/d$.  Lower
inset: estimate of $\nu_{p}$ using Eq.\ref{ro}.}
\label{fig2}
\end{figure}
We measured the microwave response at 47.7 GHz by means of a dielectric
resonator operating in the TE$_{011}$ mode.  The temperature and field
dependence of $Q$ factor and resonant frequency yielded the complex
resistivity $\rho_{1}+\mathrm{i}\Delta\rho_{2}$.  In Fig.\ref{fig2} we
report $\rho_{1}$ and $\rho_{2}(T,H)-\rho_{2}(T,0)$ at some selected
field.  It is seen that the transition is rather broad, as somehow
expected: microwaves probe the whole thickness of the film, and an
area of order $\sim$5 mm$^{2}$, so that the effect of
nonsuperconducting inclusions and of the disorder introduced shows up
as a broadening of the transition.  In the vortex state many processes
are in general responsible for the microwave response.  We found that
a large part of the dissipation was sublinear at low fields, probably
indicating GB phenomena.  At fields larger than $\sim$ 0.4 T the
complex response was linear in the applied field.  We report in the inset of
Fig.\ref{fig2}
$\nu_{p}$ as determined from the data using Eq.\ref{ro}. Very high
values $\nu_{p}\sim$50 Ghz are attained up to 77 K, indicating
extremely high vortex pinning and steep potential wells.  As expected,
the pinning frequency rapidly drops to zero as $T\rightarrow T_{c}$.
We conclude that the introduction of BZO inclusion of nanometric size
in YBCO films has improved the intragrain vortex pinning at
high microwave frequencies. While the effect of the inclusions on GB 
needs further study, it appears promising for improving
intragrain pinning.


\begin{thebibliography}{00}

\bibitem{mcmanus} J.L. MacManus Driscoll et al., {\it Nature
Materials} {\bf 3} (2004) 439

\bibitem{fabbri} F. Fabbri et al., {\it Supercond. Sci. Technol.} {\bf 
13} (2000) 1492

\bibitem{grcc} J. I. Gittleman and B. Rosenblum, {\it Phys. Rev. Lett.} 
{\bf 16}, 734 (1966);
M. W. Coffey and J. R. Clem, {\it Phys. Rev. Lett.} {\bf 
67} (1991) 386

\bibitem{tsuchiya}
Y. Tsuchiya et al., {\it Phys. Rev. B} 
{\bf 63}, 184517 (2001).

\end{thebibliography}
\end{document}